\def\lesssim{\mathrel{\hbox{\rlap{\hbox{\lower4pt\hbox{$\sim$}}}\hbox{$<$}}}}

\documentclass[10pt,a4paper,twoside]{article}
\usepackage{baltlat6}
\usepackage{epsfig}
\usepackage{lscape}
\usepackage{wrapfig}
\begin{document}
\ \ \vspace{-0.5mm}

\setcounter{page}{351}
\vspace{-2mm}

\titlehead{Baltic Astronomy, vol.\,17, 351--361, 2008}

\titleb{SIMCLUST -- A PROGRAM TO SIMULATE STAR CLUSTERS}

\begin{authorl}
\authorb{V.~Deveikis}{1},
\authorb{D.~Narbutis}{1,2},
\authorb{R.~Stonkut\.{e}}{2},
\authorb{A.~Brid\v{z}ius}{2} and
\authorb{V.~Vansevi\v{c}ius}{1,2}
\end{authorl}

\begin{addressl}
\addressb{1}{Vilnius University Observatory, \v{C}iurlionio 29,
Vilnius LT-03100, Lithuania \\ viktoras.deveikis@ff.vu.lt}
\addressb{2}{Institute of Physics, Savanori\c{u} 231, Vilnius
LT-02300, Lithuania}
\end{addressl}

\submitb{Received 2008 November 18; accepted 2008 December 30}

\begin{summary} We present a program tool, \textsc{SimClust}, designed
for Monte-Carlo modeling of star clusters. It populates the available
stellar isochrones with stars according to the initial mass function
and distributes stars randomly following the analytical surface number
density profile. The tool is aimed at simulating realistic images
of extragalactic star clusters and can be used to: (i) optimize object
detection algorithms, (ii) perform artificial cluster tests for the
analysis of star cluster surveys, and (iii) assess the stochastic
effects introduced into photometric and structural parameters of
clusters due to random distribution of luminous stars and non-uniform
interstellar extinction. By applying \textsc{SimClust}, we have
demonstrated a significant influence of stochastic effects on the
determined photometric and structural parameters of low-mass star
clusters in the M\,31 galaxy disk. The source code and examples are
available at the \textsc{SimClust} website: {\tt http://www.astro.ff.vu.lt/software/simclust/}.
\end{summary}


\resthead{A program tool to simulate star clusters}
{V.~Deveikis, D.~Narbutis, R.~Stonkut\.{e} et al.}

\sectionb{1}{INTRODUCTION}

Recent studies of star clusters have extended our understanding of
these gravitationally bound systems, which have also been successfully
used to investigate evolution of their host galaxies; see a review
by Kroupa (2008) and references therein. The recent observational
data (see review by Renzini 2008) suggest the presence of multiple
stellar populations in star clusters. Theoretical modeling (e.g.,
Lamers et al. 2006; Kruijssen \& Lamers 2008) indicates that the
stellar mass function evolves from the initial mass function (IMF)
due to dynamical effects as the cluster ages. However, the assumption
that a star cluster consists of a simple stellar population (SSP)
is still an attractive simplification in extragalactic studies.

Although due to the improved quality of observations the detection
of extensive low-mass extragalactic star cluster samples is now
possible, the straightforward application of SSP models to derive
their evolutionary parameters is valid only for high-mass clusters
(Cervi\~{n}o \& Luridiana 2004), even neglecting effects of dynamical
evolution and loss of stars. The morphology of young low-mass star
clusters also depends on the stochastic distribution of luminous
stars and non-uniform extinction, therefore, their structural, as
well as photometric parameters are biased. These stochastic uncertainties
are also very important for the study of selection effects in extragalactic
surveys of unresolved and semi-resolved star clusters, which should
be quantified in terms of age, mass, extinction and size.

Therefore, we have developed a program tool, \textsc{SimClust}, by
adopting Monte-Carlo modeling, which was previously used to study
cluster color distributions (e.g., Girardi \& Bica 1993; Bruzual
2002), but extending this approach to simulate {\it realistic} images.
Thus, proper investigation of stochastic effects and observational
errors on structural parameters and aperture CCD photometry, as well
as the effects of non-uniform extinction in a star cluster, become
now possible. The program code is available under the GNU license\footnote{~GNU
General Public License.} and may be modified to incorporate different
isochrones and stellar surface number density distribution profiles.
The source code and examples are available at the \textsc{SimClust}
website\footnote{~Download \textsc{SimClust} from \tt http://www.astro.ff.vu.lt/software/simclust/.}.

\textsc{SimClust} was applied to investigate stochastic effects and
the validity of the SSP model fitting for compact star cluster sample
in the disk of the M\,31 galaxy (Narbutis et al. 2008). In this
study we have simulated numerous images of artificial clusters of
typical age, mass and size, characteristic of the M\,31 clusters
(Vansevi\v{c}ius et al. 2009). Significant influence of stochastic
effects was found for low-mass ($\lesssim\!3000~M_{\odot}$) star
clusters younger than $\sim$\,100~Myr.

This program tool shares a similar approach to the star cluster
simulation problem investigated by Popescu \& Hanson (2008). \textsc{SimClust}
uses different stellar isochrone sets, initial mass function sampling
technique and modeling of interstellar extinction, thus the user
is provided with an independent tool to perform Monte-Carlo simulations
of extragalactic star clusters and to check the robustness of cluster
models.

We present the star cluster simulation method in \S\,2, describe
the \textsc{SimClust} program in \S\,3 and investigate stochastic
effects for cluster sample in the M\,31 galaxy in \S\,4. We also
discuss the implications of Monte-Carlo modeling in extragalactic
star cluster studies on both the estimation of star cluster sample
completeness and determination of object parameters in the stochastic
model framework.

\sectionb{2}{STAR CLUSTER SIMULATION}

\textsc{SimClust} program populates a stellar isochrone of a given
age and metallicity with stars. Randomly generated masses of stars
are weighted on the IMF and the process is repeated until a given
mass of model star cluster, $M$, is reached. Stellar magnitudes are
computed assuming a distance to the artificial ``extragalactic''
cluster, and stars are scattered in the image plane by the analytical
surface number density profile. The effect of non-uniform extinction
can be applied for individual stars. Finally, {\it realistic} multi-band
images of model cluster are rendered. We describe the details of
each step below.

\subsectionb{2.1}{Isochrones and extinction -- stellar magnitudes}

For stellar magnitudes in different photometric passbands as a function
of {\it initial}\,\footnote{~Initial and actual (smaller due to mass
loss) stellar masses are provided in isochrone tables.} stellar mass
in solar-mass unit, $m$, age, $t$, and metallicity, [M/H], we use
a new set of Padova\footnote{~Padova isochrones: \tt http://stev.oapd.inaf.it/cgi-bin/cmd.}
isochrones by Marigo et al. (2008). These isochrones are combined
of models: (i) up to early-AGB from Girardi et al. (2000), (ii)
detailed TP-AGB for $m \leq 7\,M_{\odot}$ from Marigo \& Girardi
(2007) and (iii) for $m > 7\,M_{\odot}$ from Bertelli et al. (1994),
and take into account many critical aspects of AGB evolution, important
for star cluster simulations. In this study we have used the Marigo
et al. (2008) isochrones in the Johnson-Cousins {\it UBVRIJHK}
photometric system (Ma{\'{\i}}z Apell\'{a}niz 2006; Bessell 1990).

The isochrone consists of points for a discrete stellar mass with
tabulated corresponding magnitudes for each passband. For every
randomly generated star (see \S\,2.2), we calculate its magnitude
in each passband by linear interpolation between the closest mass
points on the isochrone. The accuracy of the magnitude is directly
related to the mass sampling on the isochrone, and the linear interpolation
should have negligible effect on the accuracy of the integrated
cluster magnitude.

The interstellar extinction is applied for individual stars assuming
the standard extinction law (Cardelli et al. 1989), calculated as
a function of the given color excess, $E_{B-V}$. To investigate the
effects of non-uniform (differential) extinction in young clusters
(Yadav \& Sagar 2001), we assume for simplicity that reddening of
individual stars in the cluster is described by a Gaussian distribution
over mean $E_{B-V}$ and a standard deviation $\sigma_{E_{B-V}}$.
This distribution is generated using {\tt gsl\_ran\_gaussian} function
from the GNU Scientific Library\footnote{~GSL version 1.9: \tt
http://www.gnu.org/software/gsl/.}. Setting the $\sigma_{E_{B-V}}
= 0$, all stars are reddened by the same amount. In case of $\sigma_{E_{B-V}}
> 0$, the lower limit of reddening, corresponding to the foreground
extinction, is applied.

\subsectionb{2.2}{IMF -- stellar mass distribution}

The stellar initial mass function defines a number of stars, $dN$,
within a stellar mass range, $dm$. In this work we use the universal
IMF (Kroupa 2001), defined as the multi-power law, which has the
general form of

\begin{equation}\label{eq:imf1}
\displaystyle
\xi(m)=dN/dm=C_{\rm IMF}\cdot b_{i}\cdot m^{\alpha_{i}}\,,
\end{equation}
where $\alpha_{i}$ is the IMF slope for the mass interval $[m_{i-1},
m_{i}]$, $C_{\rm IMF}$ and $b_{i}$ are the normalization and function
continuation constants, respectively.

We follow the practical numerical formulation of the IMF by Pflamm-Altenburg
\& Kroupa (2006), and implement their shared library ``Libimf''\footnote{~Libimf:
\tt http://www.astro.uni-bonn.de/$\sim$webaiub/dwld/libimf-05-10-06.tar.gz.}
in our code; namely the function {\tt imf\_init\_multi\_power}, which
allows us to define an arbitrary number $i$ of mass intervals $[m_{i-1},
m_{i}]$, and corresponding power law slopes $\alpha_{i}$.

Because of the lowest stellar mass limit $m_{\rm min}$ = 0.15\,$M_{\odot}$
for the isochrones used, the IMF slope (Eq. 2, Kroupa 2001) transforms
into a simple equation:
\begin{equation}\label{eq:imf2}
\displaystyle
\cases{\begin{array}{l}
\alpha_{0}=-1.3, \qquad m/M_{\odot}<0.50\,; \\
\alpha_{1}=-2.3, \qquad m/M_{\odot}\geq 0.50\,.
\end{array}}
\end{equation}

Before a random number generator can be applied to sample the IMF,
it has to be normalized, i.e., the constant $C_{\rm IMF}$ has to be
defined. We follow the normalization strategy by Weidner \& Kroupa
(2004), implemented by Pflamm-Altenburg \& Kroupa (2006) in the
``Libimf'' library as a function {\tt imf\_norm\_wk04}. It requires
two additional free parameters: (i) $m_{\rm max*} = 150\,M_{\odot}$
-- the maximum physically possible stellar mass, and (ii) $m_{\rm
max}$ -- the expected maximum stellar mass in a given cluster of
mass $M$. A reasonable assumption is to define both $C_{\rm IMF}$
and $m_{\rm max}$ in such a way, that there is {\it only one} massive
star of mass $m_{\rm max}$ in the mass range $[m_{\rm max},m_{\rm
max*}]$. Thus, the expected maximum stellar mass is a function of
cluster's mass, i.e., $m_{\rm max} = m_{\rm max}(M)$. The normalization
is a required procedure for the present application of the Monte-Carlo
method, since now the IMF can be treated as a probability density
function in the range between $m$ and $m+dm$.

\begin{figure}[!t]
\vbox{\centerline{\psfig{figure=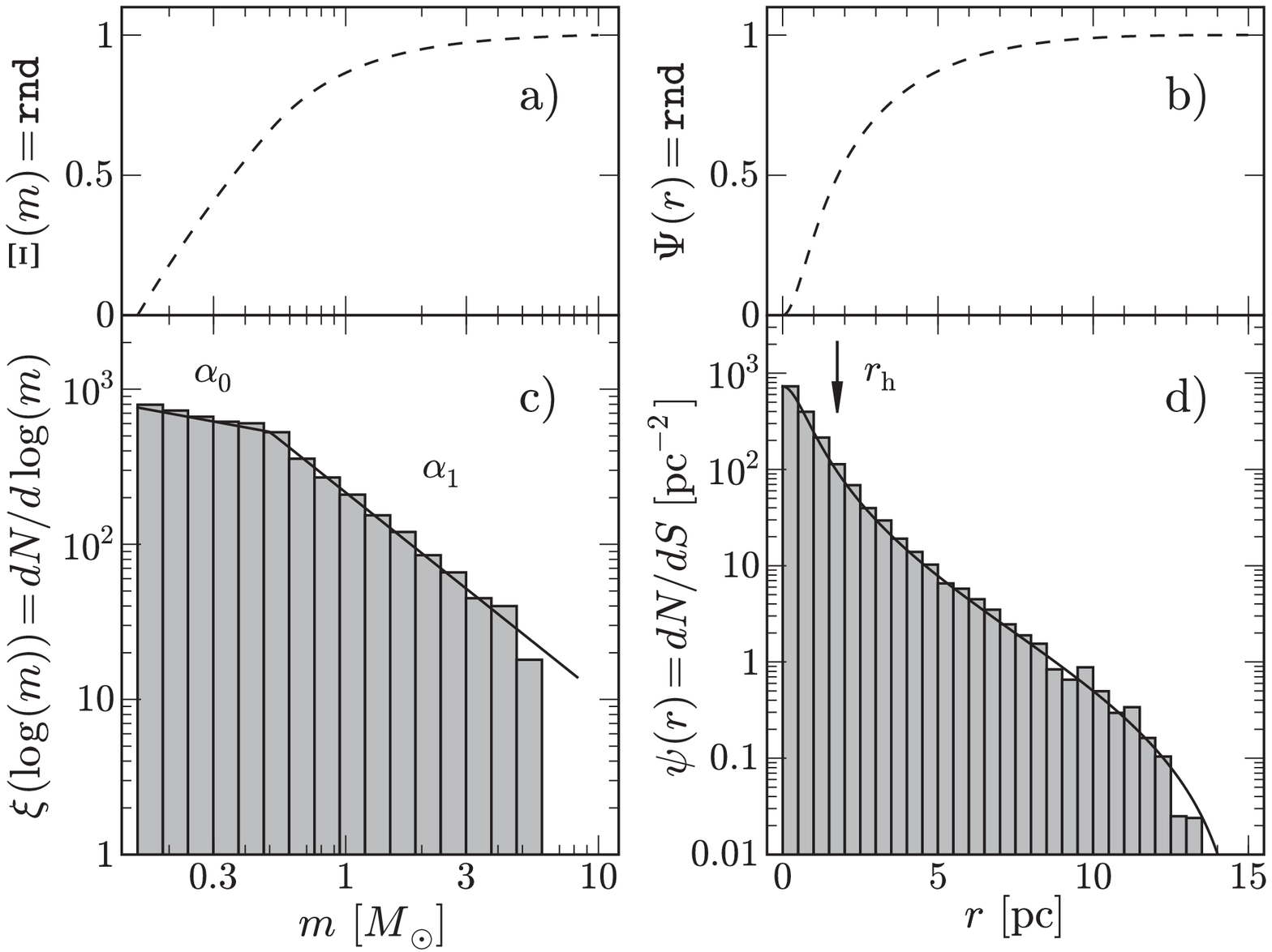,width=110truemm,angle=0,clip=}}
\vspace{0.5mm}
\captionb{1}{Properties of a model star cluster with $t=100$~Myr and
``luminous'' mass $M_{\rm lum}=3000~M_{\odot}$, consisting of $N_{\rm
lum}=5300$ stars. The cumulative mass distribution function $\Xi(m)$
of the IMF (see Eq.~\ref{eq:imf3}) and the cumulative radial distribution
function $\Psi(r)$ of the King model (see Eq.~\ref{eq:king2}), displayed
in panels (a) and (b), respectively, are used to sample mass, $m$,
and radial distance, $r$, by means of uniform random numbers, {\tt
rnd}. The stellar mass distribution, $\xi(\log\kern1pt(m))$ = $dN/d\kern1pt\log(m)$,
is displayed vs. mass, $m$, in panel (c), overplotted with slope
$\alpha_{0}=-1.3$ and $\alpha_{1}=-2.3$ lines of IMF (see Eq.~\ref{eq:imf2}).
The lower and upper mass limits are defined by the isochrone. The
surface number density of stars, $\psi(r) = dN/dS$, is plotted vs.
radial distance, $r$, in panel (d), obeying King (1962) model profile
of $r_{\rm c}=0.75$~pc and $r_{\rm t} = 15$~pc (see Eq.~\ref{eq:king1});
half-light radius $r_{\rm h} \sim 1.7$~pc is indicated by an arrow.}}
\end{figure}

The IMF sampling technique can be summarized as follows: from
Eq.~\ref{eq:imf1} a cumulative mass distribution function, giving
a probability that mass $m^{\prime} \leq m$, is constructed:
\begin{equation}\label{eq:imf3}
\displaystyle
\Xi(m)=\int\limits_{m_{\rm min}}^{m}{\xi(m^{\prime})}\,dm^{\prime}\,.
\end{equation}
This function (Eq.~\ref{eq:imf3}) is continuous between 0 and 1
(see Figure~1a). Thus, a uniform random number, ${\tt rnd}$, obtained
from long periodicity {\tt gsl\_rng\_taus} randomizer function of
the GNU Scientific Library, is attributed to $\Xi(m) = {\tt rnd}$.
The stellar mass $m$ values are sampled from the IMF via inverse
function
\begin{equation}\label{eq:imf4}
\displaystyle
m=F^{-1}\left(\Xi(m) = {\tt rnd}\right)\,,
\end{equation}
which has an analytical solution. We use the ``Libimf's'' {\tt
imf\_dice\_star\_cl} function to generate an array of stellar masses
according to the supplied IMF.

The mass of the cluster derived by the SSP model fitting and used
in population analysis is usually the initial mass at the cluster's
birth. Following this definition, we generate the initial mass function
of stars at age $t = 0$~Myr, where individual stellar masses populate
the range $[m_{\rm min},m_{\rm max}]$, as defined previously. We
denote the sum of initial stellar masses to the mass of the cluster,
$M$. When the isochrone of age $t > 0$~Myr is sampled, only stars,
which are present at that age ($m_{\rm max} = m_{\rm max}(t) < m_{\rm
max}(t = 0\,{\rm Myr})$) contribute to the luminosity of cluster,
i.e., ``luminous'' mass, $M_{\rm lum}$, is always lower than the
``initial'' mass, $M$, of the cluster. Also, the mass-loss of stars
due to stellar winds can be accounted for and directed to the output.

An example of the resulting stellar mass distribution, $\xi\kern1pt(\log\,(m))
= dN/d\kern1pt\log\,(m)$, displayed vs. mass, $m$, of $t = 100$~Myr
model star cluster of ``luminous'' mass $M_{\rm lum} = 3000~M_{\odot}$,
consisting of $N_{\rm lum} = 5300$ stars, is shown in Figure~1c,
overplotted with the IMF slope $\alpha_{0}=-1.3$ and $\alpha_{1}=-2.3$
lines (see Eq.~\ref{eq:imf2}).

\subsectionb{2.3}{Radial surface number density profile -- 2-D spatial
distribution}

Once the mass and multi-band magnitudes of a star are known, its
spatial position has to be defined. For simplicity we assume that
the ``observed'' positions of the stars are distributed randomly
in the 2-D plane, obeying a circular King (1962) model.

The King model describes the radial surface number density profile,
i.e., the number of stars, $dN$, per unit area, $dS$ and is defined
by the central surface number density, $\psi_{0}$, the core radius,
$r_{\rm c}$, and the tidal radius, $r_{\rm t}$:
\begin{equation}\label{eq:king1}
\displaystyle
\psi\,(r) = dN/dS = dN/2\pi r\,dr = \psi_{0}\left[{\left({1+
\frac{{r^{2}}}{{r_{\rm c}^{2}}}}\right)^{-1/2}-\left({1+\frac{{r_{\rm
t}^{2}}}{{r_{\rm c}^{2}}}}\right)^{-1/2}}\right]^{2}\,,
\end{equation}
where $r$ is the distance from cluster's center.

The cumulative radial distribution function of the circular King
model giving probability that the distance $r^{\prime}\leq r$, is
defined by formula:
\begin{eqnarray}\label{eq:king2}
\displaystyle
\lefteqn{\Psi(r)=2\pi\,C_{\rm King}\int\limits_{0}^{r}r^{\prime}
\psi(r^{\prime})\,dr^{\prime} = } \nonumber \\ &
\pi\,C_{\rm King}\cdot r_{\rm c}^{2}\left[\ln\left(1+(r/r_{\rm
c})^{2}\right)-4\left(\sqrt{(r_{\rm c}^{2}+r^{2})/\lambda}-\sqrt{r_{\rm
c}^{2}/\lambda}\right)+r^{2}/\lambda\right]\,,
\end{eqnarray}
where $\lambda=r_{\rm c}^{2}+r_{\rm t}^{2}$, and $C_{\rm King}$ is
the normalization constant, defined as
\begin{equation}\label{eq:king3}
\displaystyle
C_{\rm King}=\left(2\pi\int\limits_{0}^{r_{\rm t}}r^{\prime}\psi(r^{\prime})\,
dr^{\prime}\right)^{-1}\,.
\end{equation}

The $\Psi\,(r)$ function (Eq.~\ref{eq:king2}) is continuous between
0 and 1 (see Figure~1b). Thus, a uniform random number, ${\tt rnd}$,
is attributed to $\Psi\,(r) = {\tt rnd}$ to sample the distance $r$
values from the King model via inverse function
\begin{equation}\label{eq:king4}
\displaystyle
r=F^{-1}\left(\Psi(r) = {\tt rnd}\right)\,,
\end{equation}
which does not have an analytical expression. Since the function
$\Psi(r)$ is monotonously increasing, therefore, for a generated
random number, ${\tt rnd}$, we can solve Eq.~\ref{eq:king4} via
Eq.~\ref{eq:king2}, iteratively by the bisection method until the
solution of $r$ converges.

An example of the resulting surface number density of stars, $\psi(r)
= dN/dS$, displayed vs. radial distance, $r$, of the model star
cluster, consisting of $N_{\rm lum} = 5300$ stars, is shown in Figure~1d,
overplotted with the King (1962) model profile of $r_{\rm c}=0.75$~pc
and $r_{\rm t}=15$~pc (see Eq.~\ref{eq:king1}); half-light\footnote{~In
this case this is the radius containing the half number of the stars.}
radius $r_{\rm h} \sim 1.7$~pc is indicated by an arrow.

Finally, the azimuth angle, $\varphi$, is assigned randomly for a
star. Stellar positions in parsecs ($x,y$) are computed in respect
to the cluster center:
\begin{equation}\label{eq:position}
\cases{\begin{array}{l}
\displaystyle
\varphi=2\pi\cdot{\tt rnd}\,,\\
x=r\cdot\cos(\varphi)\,,\\
y=r\cdot\sin(\varphi)\,.\\
\end{array}}
\end{equation}

\subsectionb{2.4}{Realistic images}

The distance to the star cluster, $D$, and the photometric zero-point
are used to transform the absolute magnitudes of stars to the instrumental
magnitude system in each photometric passband. Stellar positions
in the image are computed assuming the provided pixel scale, $\theta$,
and the center coordinates of star cluster.

Finally, these data are passed to the external ``SkyMaker''\footnote{~SkyMaker version 3.1.0:
\tt http://terapix.iap.fr/soft/skymaker.} program, which renders
{\it realistic} images for each photometric passband, assuming the
provided point-spread function (PSF). Usually the PSF is derived
from the real image by using the DAOPHOT program package (Stetson
1987) implemented in the IRAF program system (Tody 1993). Since
``SkyMaker'' has many image rendering options, they can be used to
create realistic image noise properties, or even simulate images
for an arbitrary telescope configuration or observational conditions.

\sectionb{3}{\textsc{SimClust} -- PROGRAM DESCRIPTION}

Here we present a short description of \textsc{SimClust} usage, the
properties of input files, parameter configuration and output files.
The schematic representation of the data flow is shown by means of
a block diagram given in Figure~2.

\begin{figure}[!t]
\vbox{\centerline{\psfig{figure=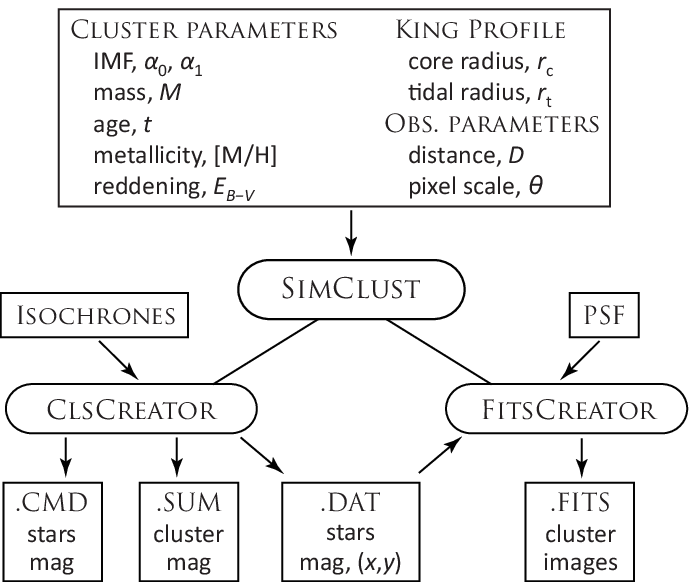,width=85truemm,angle=0,clip=}}
\vspace{2mm}
\captionb{2}{Schematic representation of \textsc{SimClust}, which
consists of two sub-programs -- \textsc{ClsCreator} and \textsc{FitsCreator}.
Arrows indicate the parameter and data flow.}}
\end{figure}
\vskip3mm

\textsc{SimClust} is written in the Perl and C++ languages and can
be compiled under the ``UNIX/Linux'' systems. Installation script
{\tt compile} checks that external program ``SkyMaker'' and all
necessary libraries are available on the computer and builds executables.
The configuration file {\tt example}, which contains main parameters
of the model star cluster and instructions on how the images should
be simulated, is passed to {\tt SimClust} which controls two sub-programs,
\textsc{ClsCreator} and \textsc{FitsCreator}.

The \textsc{ClsCreator}, which randomly populates the input stellar
isochrones, has to be executed first. There are three possible output
options: (i) {\tt .cmd} file, containing magnitudes of individual
stars, (ii) {\tt .sum} -- integrated magnitudes of star clusters,
and (iii) {\tt .dat} -- stellar magnitudes and positions. Later the
\textsc{FitsCreator} can be asked to pass {\tt .dat} files and PSF
to the ``SkyMaker'' to simulate cluster images ({\tt .fits}).

The detailed description of each configuration parameter is documented
in the help file and illustrated with examples.

\sectionb{4}{STOCHASTIC EFFECTS OF STAR CLUSTERS}

The accuracy of cluster age and mass from broad-band photometry,
derived by applying traditional SSP models, as well as the accuracy
of structural cluster parameters, depend on the clusters age, mass
and size, image spatial resolution, data analysis techniques and
stochastic effects. To account properly for stochastic effects, we
have developed \textsc{SimClust} and applied it to investigate properties
of star clusters in the extensively studied South-West field of the
M\,31 galaxy (Kodaira et al. 2004, 2008; Narbutis et al. 2006, 2007b,
2008). Previously, for this cluster sample the influence of individual
semi-resolved bright stars on the accuracy of the structural cluster
parameters was discussed by \v{S}ablevi\v{c}i\={u}t\.{e} et al. (2006,
2007), and the influence on the accuracy of aperture photometry was
studied by Narbutis et al. (2007a).

In the following we provide an example of \textsc{SimClust} application
to study star clusters in M\,31. We describe the cluster simulations
and the expected photometric and half-light radii uncertainties,
derived for typical star clusters of our sample. For the present
study we assume the typical parameter values of compact star clusters
(Vansevi\v{c}ius et al. 2009): age $\sim$\,100~Myr, mass $\sim$\,3000\,$M_{\odot}$
and half-light radius $\sim$\,1.5~pc.

\subsectionb{4.1}{Star cluster model setup}

Since the majority of objects in the M\,31 sample are young low-mass
star clusters, we were motivated to investigate the influence of
stochastic distribution of bright stars on both photometric and
structural properties. For this purpose, we have chosen parameters
adequate to the characteristic values of our sample: (i) five age
cases, $t=20$, 100, 500~Myr, 2.5 and 12.5~Gyr; (ii) ``luminous''
masses, which correspond to the adopted ages, $\log\kern1pt(M_{\rm
lum}/M_{\odot})$ = 3.5, 3.5, 3.5, 4.0 and 5.0; (iii) metallicity,
[M/H] = --\,0.4 dex; (iv) reddening, $E_{B-V}$ = 0.25; (v) distribution
of stars according to the circular King model of $r_{\rm c}=0.75$~pc
and $r_{\rm t}$ = 15~pc, i.e., $r_{\rm h}$\,$\approx$\,1.7~pc. A
thousand model clusters were produced for each age-mass cases.

Model star clusters were placed at the distance of M\,31, $D=785$~kpc,
derived by McConnachie et al. (2005); thus 1\arcsec\ = 3.8~pc. For
simplicity, zero photometric background and noise was assumed for
image simulation.

\begin{figure}[!t]
\vbox{\centerline{\psfig{figure=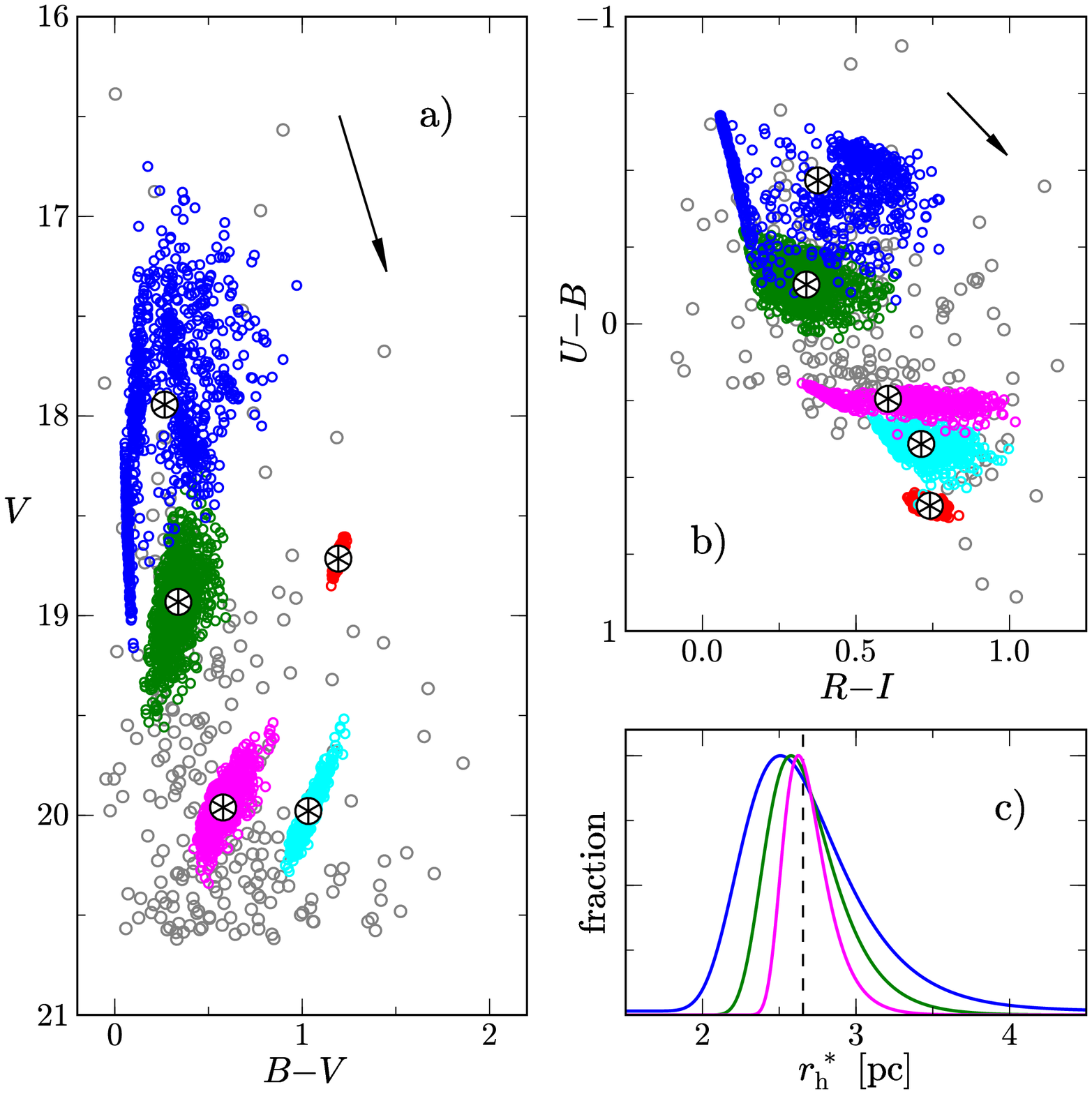,width=105truemm,angle=0,clip=}}
\vspace{0.5mm}
\captionb{3}{Modeling of stochastic effects. Panel (a). The diagram
$V$ vs. $B$--$V$ of 238 high probability star cluster candidates in
M\,31 (gray circles) from Vansevi\v{c}ius et al. (2009), overplotted
with five cases of model clusters for the ages $t$ = 20, 100, 500~Myr,
2.5 and 12.5~Gyr, and ``luminous'' masses $\log\,(M_{\rm lum}/M_{\odot})$
= 3.5, 3.5, 3.5, 4.0 and 5.0 (blue, green, magenta, cyan and red
symbols), 1000 models for each case. Asterisks indicate the position
of the SSP model, vector indicates interstellar reddening of $E_{B-V}$
= 0.25. Panel (b). The diagram $U$--$B$ vs. $R$--$I$, symbols are the
same as in (a). Panel (c). Distributions of model clusters half-light
radii, measured using a curve-of-growth method, for the ages $t$ =
20, 100 and 500~Myr; the dashed line indicates the true $r_{\rm h}^{*}$
= 2.6~pc.}}
\end{figure}

\subsectionb{4.2}{Photometric properties}

To examine the photometric properties of model clusters, images
equivalent to the Local Group Galaxies Survey data (Massey et al.
2006) with homogenized PSF of ${\rm FWHM} \approx 1.5$\arcsec\ were
simulated; image scale $\theta$ = 0.27\arcsec\ per pixel. The {\it
UBVRI} photometry was performed with IRAF's \verb"phot" through
apertures of 3\arcsec\ in diameter -- a typical size used for the
sample clusters (see Narbutis et al. 2008 for details of photometry).

The resulting diagrams $V$ vs. $B$--$V$ and $U$--$B$ vs. $R$--$I$
are displayed in Figures~3a and 3b, respectively, overplotted on
238 high probability cluster candidates, selected by Vansevi\v{c}ius
et al. (2009). The stochastic scatter of model luminosity and color
dominates over the typical photometric errors. The indicated SSP
model points\footnote{~Padova SSP models: \tt http://stev.oapd.inaf.it/cgi-bin/cmd.}
(Marigo et al. 2008) of [M/H] = --\,0.4 dex, reddened by $E_{B-V}$
= 0.25, show that SSP colors at $t$ = 20~Myr and 500~Myr do not
represent model cluster colors and are disturbed asymmetrically
around the SSP position in two-color diagram due to stochastic existence
of bright blue and red supergiants, RGB and AGB stars, prominent
at these ages. However, for the remaining three model ages, the
centers of color distribution coincide well with the SSP models.
The color distribution of sample clusters is covered by the extent
of stochastic models with assumed typical reddening of $E_{B-V}$ =
0.25. We note, that the scatter of colors of $t$ = 100~Myr and
$\log\,(M_{\rm lum}/M_{\odot})$ = 3.5 models resemble that of real
objects located around [0.4,$-0.2$] in Figure~3b.

\subsectionb{4.3}{Structural properties}

To examine the structural properties of model clusters, images equivalent
to the Suprime-Cam $V$-band data of FWHM$_{\rm PSF} \approx 0.7$\arcsec\
were simulated; the image scale $\theta$ = 0.20\arcsec\ per pixel
(see Narbutis et al. 2008 for Suprime-Cam survey data details).

IRAF's \verb"phot" was applied to obtain the profile of the enclosed
flux of a cluster model using the curve-of-growth method, yielding
the half-light radius, $r_{\rm h}^{*}$. Note, however, that observational
effects (PSF) were not taken into account.

The derived half-light radii $r_{\rm h}^{*}$ data histograms of
$10^{3}$ models for ages $t=20$, 100 and 500~Myr were constructed.
After smoothing, they are displayed in Figure~3c. The scatter around
the true $r_{\rm h}^{*} \approx 2.6$~pc value is due to a stochastic
distribution of bright stars in model star clusters. The stochastic
effect is most significant for the $t=20$~Myr models, and decreases
to older and more massive clusters. The values of $r_{\rm h}^{*}$
are smaller for clusters, which have bright stars residing in the
cluster centers, dominating the overall luminosity. Contrary, when
bright stars are located far from the center, the derived $r_{\rm
h}^{*}$ values are increased. The standard deviation of $r_{\rm
h}^{*}$ distribution of the youngest model clusters ($t=20$~Myr)
is $\sim$\,0.5~pc and decreases to $\sim$\,0.1~pc for the oldest
ones ($t=500$~Myr).

\sectionb{5}{DISCUSSION AND CONCLUSIONS}

Simulations of star cluster images are becoming important to evaluate
the performance of analysis techniques of observational data, as
have been shown by, e.g., Ascenso et al. (2008) for the case of mass
segregation in star clusters. Popescu \& Hanson (2008) emphasize
the importance of realistic star cluster simulation with incorporation
of interstellar extinction to answer the question about the lack or
existence of massive star clusters in galaxies. The current knowledge
about the most massive clusters can be limited due to unknown selection
effects.

There is no technique available (Cervi\~{n}o \& Luridiana 2006) to
properly account for stochastic effects when photometric data are
interpreted in the framework of SSP models, except of star cluster
simulation and assessment of the reliability of the SSP model performance
in certain age and mass domains. It might turn-out that the interpretation
of properties of the cluster population, especially in the young
low-mass domain, will change dramatically once the stochastic effects
are taken into account. On the other hand, the fundamental limitation
on the amount of information, which can be extracted from integrated
photometry alone, will be established and other techniques shall be
developed.

It should be emphasized here, that even for the high-mass star clusters,
where stochastic uncertainties are comparable to the observation
photometric errors, parameter degeneracies in broad-band photometry
are unavoidable; see Brid\v{z}ius et al. (2008) for the performance
analysis of the {\it UBVRIJHK} photometric system.

The tests performed in this study show that the stochastic effects
are significant for low-mass young clusters. The age degeneracy due
to stochastic color scatter is observed in all model cases. The
reddening vector in Figure~3 implies that stochastic color shift
could also mimic reddening. Therefore, when SSP model fitting is
applied, the derived cluster mass becomes more uncertain, not only
due to the miss-match of age and extinction, but also due to the
intrinsic scatter of luminosity. Visual inspection of the model
cluster color images shows a close resemblance to the real compact
cluster candidates in respect to their semi-resolved appearance and
stochastic emergence of bright blue or red stars.

The morphology of low-mass star clusters is strongly dependent on
the stochastic distribution of luminous stars and the spatial resolution
of the image. A usual ``by-eye'' selection technique leads to systematic
biases in the magnitude limited samples, thus automatic detection
should be used instead. To properly estimate the completeness of
cluster sample, the artificial star cluster tests should be performed
using {\it realistic} simulations of cluster images on individual
star basis, instead of just assuming the smooth surface brightness
distribution profiles.

The stochastic SSP model framework, which could replace the traditional
SSP model fitting to quantify ages, masses and extinctions of star
clusters from their integrated broad-band photometry, is still to
be developed. The \textsc{SimClust} program tool may serve as a basis
for this method, as well as for extensive extragalactic star cluster
studies.

\thanks{We thank A. Ku\v{c}inskas for critical comments and numerous
suggestions, which helped to improve the paper. This work was financially
supported in part by a Grant of the Lithuanian State Science and
Studies Foundation.}

\References

\refb Ascenso~J., Alves~J., Lago~M.\,T.\,V.\,T. 2008, arXiv:0811.3213

\refb Bertelli~G., Bressan~A., Chiosi~C., Fagotto~F., Nasi~E. 1994,
A\&AS, 106, 275

\refb Bessell~M.~S. 1990, PASP, 102, 1181

\newpage

\refb Brid\v{z}ius~A., Narbutis~D., Stonkut\.{e}~R., Deveikis~V.,
Vansevi\v{c}ius~V. 2008, Baltic Astronomy, 17, 337

\refb Bruzual~A.~G. 2002, in {\it Extragalactic Star Clusters}, IAU
Symp. 207, Eds. D. Geisler et al., ASP, p.\,616

\refb Cardelli~J.~A., Clayton~G.~C., Mathis~J.~S. 1989, ApJ, 345, 245

\refb Cervi\~{n}o~M., Luridiana~V. 2004, A\&A, 413, 145

\refb Cervi\~{n}o~M., Luridiana~V. 2006, A\&A, 451, 475

\refb Girardi~L., Bica~E. 1993, A\&A, 274, 279

\refb Girardi~L., Bressan~A., Bertelli~G., Chiosi~C. 2000, A\&AS, 141,
371

\refb King~I. 1962, AJ, 67, 471

\refb Kodaira~K., Vansevi\v{c}ius~V., Brid\v{z}ius~A., Komiyama~Y.,
Miyazaki~S., Stonkut\.{e}~R., \v{S}ablevi\v{c}i\={u}t\.{e}~I.,
Narbutis~D. 2004, PASJ, 56, 1025

\refb Kodaira~K., Vansevi\v{c}ius~V., Stonkut\.{e}~R., Narbutis~D.,
Brid\v{z}ius~A. 2008, in {\it Pano\-ramic Views of Galaxy Formation
and Evolution}, ASP Conf. Ser., 399, 431

\refb Kroupa~P. 2001, MNRAS, 322, 231

\refb Kroupa~P. 2008, arXiv:0810.4143

\refb Kruijssen~J.\,M.\,D., Lamers~H.\,J.\,G.\,L.\,M. 2008, A\&A,
490, 151

\refb Lamers~H.\,J.\,G.\,L.\,M., Anders~P., de Grijs~R. 2006, A\&A,
452, 131

\refb McConnachie~A.~W., Irwin~M.~J., Ferguson~A.\,M.\,N., Ibata~R.~A.,
Lewis~G.~F., Tanvir~N. 2005, MNRAS, 356, 979

\refb Ma{\'{\i}}z Apell\'{a}niz~J. 2006, AJ, 131, 1184

\refb Marigo~P., Girardi~L. 2007, A\&A, 469, 239

\refb Marigo~P., Girardi~L., Bressan~A., Groenewegen~M.~A.~T., Silva~L.,
Granato~G.~L. 2008, A\&A, 482, 883

\refb Massey~P., Olsen~K.\,A.\,G., Hodge~P.~W., Strong~S. B.,
Jacoby~G.~H., Schlingman~W., Smith~R.~C. 2006, AJ, 131, 2478

\refb Narbutis~D., Vansevi\v{c}ius~V., Kodaira~K., \v{S}ablevi\v{c}i\={u}t\.{e}~I.,
Stonkut\.{e}~R., Brid\v{z}ius, A. 2006, Baltic Astronomy, 15, 461

\refb Narbutis~D., Vansevi\v{c}ius~V., Kodaira~K., Brid\v{z}ius~A.,
Stonkut\.{e}, R. 2007a, Baltic Astronomy, 16, 409

\refb Narbutis~D., Brid\v{z}ius~A., Stonkut\.{e}~R., Vansevi\v{c}ius~V.
2007b, Baltic Astronomy, 16, 421

\refb Narbutis~D., Vansevi\v{c}ius~V., Kodaira~K., Brid\v{z}ius~A.,
Stonkut\.{e}~R. 2008, ApJS, 177, 174

\refb Pflamm-Altenburg~J., Kroupa~P. 2006, MNRAS, 373, 295

\refb Popescu~B., Hanson~M.~M. 2008, arXiv:0811.4210

\refb Renzini~A. 2008, MNRAS, 391, 354

\refb Stetson~P.~B. 1987, PASP, 99, 191

\refb \v{S}ablevi\v{c}i\={u}t\.{e}~I., Vansevi\v{c}ius~V., Kodaira~K.,
Narbutis~D., Stonkut\.{e}~R., Brid\v{z}ius~A. 2006, Baltic Astronomy,
15, 547

\refb \v{S}ablevi\v{c}i\={u}t\.{e}~I., Vansevi\v{c}ius~V., Kodaira~K.,
Narbutis~D., Stonkut\.{e}~R., Brid\v{z}ius~A. 2007, Baltic Astronomy,
16, 397

\refb Tody~D. 1993, {\it Astronomical Data Analysis Software and
Systems II}, 52, 173

\refb Vansevi\v{c}ius~V., Kodaira~K., Narbutis~D., Stonkut\.{e}~R.,
Brid\v{z}ius~A., Deveikis~V., Semionov~D. 2009, ApJ, submitted

\refb Weidner~C., Kroupa~P. 2004, MNRAS, 348, 187

\refb Yadav~R.\,K.\,S., Sagar~R. 2001, MNRAS, 328, 370

\end{document}